\documentclass[conference,letterpaper]{IEEEtran}
\usepackage[letterpaper, left=0.63in, right=0.64in, top=0.75in, bottom=1.1in]{geometry}

\IEEEoverridecommandlockouts
\usepackage{cite}
\usepackage[linesnumbered,vlined,ruled]{algorithm2e}
\usepackage{balance}
\usepackage{amsmath,amssymb,amsfonts}
\usepackage{mathtools,breqn}
\usepackage{graphicx}
\usepackage{textcomp}
\usepackage{xcolor}
\usepackage{hyperref}
\usepackage{enumerate}
\usepackage{multicol,multirow,booktabs,tabularx}
\usepackage{caption,subcaption}
\usepackage{upgreek}
\usepackage{dblfloatfix}
\newcommand{\bigquotes}[1]{``#1''}
\newcommand{\smallquotes}[1]

\newcommand{\shim}[1]{\textcolor{black}{#1}}
\newcommand{\beom}[1]{\textcolor{black}{#1}}

\SetCommentSty{mycommfont}

\SetKwProg{Pn}{Function}{:}{\KwRet}
\SetKwInOut{Input}{Input}
\SetKwInOut{Output}{Output}
\SetKwInOut{Procedure}{Procedure}
\SetKwFunction{TEST}{TEST}

\def\BibTeX{{\rm B\kern-.05em{\sc i\kern-.025em b}\kern-.08em
    T\kern-.1667em\lower.7ex\hbox{E}\kern-.125emX}}
\begin{document}

% \title{Label Partition Selection\\for Scalable Multi-Object Tracking
% \title{Soil Property Estimation via\\Generalized Labeled Multi-Bernoulli Filtering
% \title{Clay Property Estimation in Geotechnical Databases via Labeled Random Finite Sets\thanks{This research was supported by the Defence Science Centre, an initiative of the State Government of Western Australia.}
% \title{Geotechnical Property Estimation via\\Generalized Labeled Multi-Bernoulli Filtering\thanks{This research was supported by the Defence Science Centre, an initiative of the State Government of Western Australia.}
\title{\shim{Property Estimation in Geotechnical Databases\\Using Labeled Random Finite Sets}\thanks{}\thanks{This research was supported by the Defence Science Centre, an initiative of the State Government of Western Australia.}
% \title{\ing{Geotechnical Property Estimation via\\Labeled Random Finite Sets}\thanks{This research was supported by the Defence Science Centre, an initiative of the State Government of Western Australia.}
% \title{Clay Property Estimation in Geotechnical Databases via Generalized Labeled Multi-Bernoulli Filtering\thanks{This research was supported by the Defence Science Centre, an initiative of the State Government of Western Australia.}

% \title{An~Efficient~Method~for~Geotechnical~Site~Retrieval\vspace{-1.0cm}\thanks{This research was supported by the Defence Science Centre, an initiative of the State Government of Western Australia.}
% \thanks{$\dagger$ Corresponding author}
% \\ \phantom{A}$\ast$ Corresponding author}
}
\author{
    % \IEEEauthorblockN{Changbeom Shim, Ji Youn Lee, and Diluka Moratuwage}
    % \IEEEauthorblockA{
    % \textit{School of Electrical Engineering, Computing and Mathematical Sciences}, Curtin University, Perth, Australia\\
    % changbeom.shim@curtin.edu.au \, jiyoun.lee2@postgrad.curtin.edu.au \, diluka.moratuwage@curtin.edu.au}
% \textit{$^{\dagger}$School of ???}, ??? University, ???, ??? \\
    % \IEEEauthorblockN{Changbeom Shim, Ji Youn Lee, Diluka Moratuwage, and Yon Dohn Chung$^{\ast}$}
    % \IEEEauthorblockA{
    % \textit{School of Electrical Engineering, Computing and Mathematical Sciences}, Curtin University, Perth, Australia\\
    % \textit{Department of Computer Science and Engineering}, Korea University, Seoul, South Korea\\
    % \{changbeom.shim, diluka.moratuwage\}@curtin.edu.au, jiyoun.lee2@postgrad.curtin.edu.au, ydchung@korea.ac.kr}
% \textit{$^{\dagger}$School of ???}, ??? University, ???, ??? \\
    % \IEEEauthorblockN{Changbeom Shim$^{\ast}$, Ji Youn Lee$^{\ast}$, Diluka Moratuwage$^{\ast}$, Du Yong Kim$^{\dagger}$, and Yon Dohn Chung$^{\mathsection\star}$}
    \IEEEauthorblockN{Changbeom Shim$^{1}$, Youngho Kim$^{2}$, and Craig Butterworth$^{2}$}
    \IEEEauthorblockA{
    $^{1}$\textit{School of Electrical Engineering, Computing and Mathematical Sciences}, Curtin University, Perth, Australia\\
    $^{2}$CMW Geosciences, Perth, Australia\\
    % $^{\mathsection}$\textit{Department of Computer Science and Engineering}, Korea University, Seoul, South Korea\\
    changbeom.shim@curtin.edu.au \,\:\, younghok@cmwgeo.com \,\:\, craigb@cmwgeo.com}
}

\maketitle

\begin{abstract}
    \shim{The sufficiency of accurate data is a core element in data-centric geotechnics. However, geotechnical datasets are essentially uncertain, whereupon engineers have difficulty with obtaining precise information for making decisions. This challenge is more apparent when the performance of data-driven technologies solely relies on imperfect databases or even when it is sometimes difficult to investigate sites physically. This paper introduces geotechnical property estimation from noisy and incomplete data within the labeled random finite set (LRFS) framework. We leverage the ability of the generalized labeled multi-Bernoulli (GLMB) filter, a fundamental solution for multi-object estimation, to deal with measurement uncertainties from a Bayesian perspective. In particular, this work focuses on the similarity between LRFSs and big indirect data (BID) in geotechnics as those characteristics resemble each other, which enables GLMB filtering to be exploited potentially for data-centric geotechnical engineering. Experiments for numerical study are conducted to evaluate the proposed method using a public clay database.}
\end{abstract}


%%%%%%%%%%%%%% Introduction
\section{\shim{Introduction}}\label{s:intro}
Data-centric approaches in geotechnical engineering have gained much attention along with the development of geosensing and monitoring devices \cite{national2006geological,phoon2023geotechnical}. In  \cite{phoon2023future}, three critical elements were introduced for data-centric geotechnics: i)~\textit{data centricity},  ii)~\textit{fit for (and transform) practice}, and iii)~\textit{geotechnical context}. Especially, real-world databases play a major role in achieving ``data centricity'', but geotechncial big data are known as big indirect data (BID) due to the indirect relevance to one specific site/project.  Further, typical datasets in this field are imperfect \cite{phoon2019managing}. Notwithstanding, data-driven methods remain active as an application of data-centric geotechnics. For instance, a novel machine learning technique was devised recently for shield attitude prediction \cite{wang2024data}, and data-driven site characterization (DDSC) was also elaborated in \cite{phoon2022challenges} with specific challenges, \textit{ugly data}, \textit{site recognition}, and \textit{stratification}.

Despite the significance of geotechnical data, existing knowledge and physics cannot be ignored for keeping ``geotechnical context'' \cite{phoon2022unpacking}. In this regard, a hybrid model \cite{pei2023applying} for slope stability prediction was proposed by integrating data-driven methods with domain knowledge. Later, physics guidance is additionally emphasized for the improvement of landslide susceptibility mapping \cite{pei2024landslide}. Furthermore, Bayesian approaches have been widely used in geotechnics with BID. In \cite{ching2017characterizing}, the statistical uncertainties of the site-specific trend are characterized via Bayesian learning. Moreover, hierarchical Bayesian models are particularly studied in diverse problems including the construction of quasi-site-specific probability distribution \cite{ching2021constructing}, site retrieval \cite{sharma2022hierarchical}, and the prediction of soil property \cite{wu2022quasi}, wall deflection \cite{tabarroki2024data}, and small-strain stiffness of sand \cite{tao2023hierarchical}.

The Bayesian method is a proven approach to other fields as well such as artificial intelligence, statistics, and biology. In statistical \beom{inference}, multi-object tracking (MOT), the estimation of object trajectories from noisy sensor data, has received considerable attentions. The major difficulties in the problem include unknown object numbers and measurement uncertainties at each time. Among practical MOT methods, a labeled random finite set (LRFS) solution \cite{vo2013labeled} has established its reputation with continuous \beom{technical} improvement. LRFS methods are mathematically rigorous and pragmatic in various applications, whereupon generalized labeled multi-Bernoulli (GLMB) \cite{vo2014labeled} and labeled multi-Bernoulli \cite{reuter2014labeled} filters are widely studied. The LRFS can be any types of meaning or information\beom{, e.g., multiple objects,} depending on the context, although the LRFS approach has attracted researchers and engineers in the MOT community\beom{\cite{nguyen2021biological,ishtiaq2023interaction,van2024multi,klupacs2024enhancing,van2024visual}}. 

% Estimating object trajectories from sensor data is the goal of multiple object tracking \cite{blackman1999designtracking,mahler2003multitargetPHD,mahler2007statistical,bar2011tracking}. For Multi-Object Tracking (MOT), extra difficulties exist in intrinsic noises as well as measurement uncertainties such as false alarms, missed detections, and data associations. These obstacles can be addressed by established approaches such as Multiple Hypothesis Tracking (MHT) \cite{blackman1999designtracking}, Joint Probabilistic Data Association (JPDA) \cite{bar2011tracking}, and Random Finite Set (RFS) \cite{mahler2007phdCPHD}. Thanks to the rigorous mathematical foundation, RFS-based methods have been well developed with sound theoretical and practical justifications. These include the  Probability Hypothesis Density (PHD) \cite{mahler2003multitargetPHD}, Cardinalized PHD (CPHD) \cite{mahler2007phdCPHD}, Multi-Bernoulli \cite{vo2008cardinality,vo2010joint}, and Generalized Labeled Multi-Bernoulli (GLMB) filters \cite{vo2013labeled,vo2014labeled}.

\shim{This work introduces geotechnical property estimation via LRFSs within the Bayesian framework. Due to mutual charateristics of BID and LRFSs, geotechnical parameters can be seamlessly handled by LRFS approaches. We treat each data property as each object represented by LRFSs for MOT so that the GLMB filter deals with BID. To the best of knowledge, this is the first work for geotechnical property estimation exploiting GLMB filtering. We conduct experiments using a real geotechnical database to validate the performance of the proposed approach and other potential applications in geotechnical engineering.}

\shim{The rest of this paper is composed of as follows. Section~\ref{s:gdb} introduces geotechnical databases, and Section~\ref{s:GLMB} describes GLMB filtering for geotechnical property estimation. The performance of our methodology is evaluated in Section~\ref{s:exp}. Finally, we conclude this work with future directions in Section~\ref{s:conclusion}.} 

%%%%%%%%%%%%%% Databases
\section{\shim{Geotechnical Databases}}\label{s:gdb}
\shim{Geotechnical databases are essential for dealing with diverse problems in geotechnics \cite{sharma2023spectral}. To improve relevant technologies and knowledge, the Engineering Practice of Risk Assessment and Management Committee of the International Society of Soil Mechanics and Geotechnical Engineering (ISSMGE) provides geotechnical databases (304dB)\footnote{http://140.112.12.21/issmge/tc304.htm?=6} including cone penetration test, soil or rock properties, and geospatial data. These databases are not directly related to one specific project or site, i.e., BID, but involve generic data from multiple sites or countries. For example, a generic clay database, CLAY/10/7490 \cite{ching2014transformations}, contains 7490 records from 30 countries and the following 10 clay properties are included:
\begin{enumerate}[$\cdot$]
    \vspace{0.1cm}
    \item $Y_1$ = LL (liquid limit),
    \item $Y_2$ = PI (plasticity index),
    \item $Y_3$ = LI (liquidity index),
    \item $Y_4$ = $\sigma'_v / P_a$ (normalized vertical effective stress),
    \item $Y_5$ = $S_t$ (sensitivity),
    \item $Y_6$ = $B_q$ (pore pressure ratio),
    \item $Y_7$ = $\sigma'_p / P_a$ (normalized preconsolidation stress),
    \item $Y_8$ = $s_u/\sigma'_v$ (undrained strength ratio),
    \item $Y_9$ = $(q_t-\sigma_v)/\sigma'_v$ (normalized cone tip resistance),
    \item $Y_{10}$ = $(q_t - u_2) / \sigma'_v$ (effective cone tip resistance),
    \vspace{0.1cm}
\end{enumerate}
where $\sigma_v'$ denotes vertical effective stress, $P_a$ is atmospheric pressure = 101.3 kPa, $\sigma_p'$ means preconsolidation stress, $s_u$ is undrained shear strength, $q_t$ presents (corrected) con tip resistance, and $u_2$ is pore pressure behind cone.
}

\shim{Data-driven methods in geotechnical engineering have become known for a special subcategory of data-centric geotechnics \cite{phoon2022site}. In \cite{phoon2022challenges}, DDSC refers to any site characterization methods which depend on measured data, site-specific data and existing data, neighboring sites, or beyond. However, site investigation data are generally ``MUSIC-X'' (Multivariate, Uncertain and Unique, Sparse, Incomplete, and potentially Corrupted, with ``X'' meaning the spatio-temporal dimension) \cite{phoon2019managing}. Arguably, it is necessary to devise a novel method for dealing with the characteristics of geotechnical property databases so that engineers can make decisions based on accurate site-specific data.
}

% \newpage

% % The ``M'' for ``Multivariate'' is analogous to multiple objects or sensors. ..... The ``U'' of ``Uncertain and Unique'' ... .``S''  ``Sparse'' ``I'' ``Incomplete''

% Note that the difference between ``similar database records'' and ``similar database sites'' are obvious.  .... the decision making process for geotechnical engineers .... ....  \ing{ cite this \cite{ching2023detection}} =========

% What is ``site-grouping'' information? spatial variability? - horizontal... vertical ....   ..... with increasing sparsity and degree of incompletness..... ... .... The reliability and XX of geotechnical structuring highly rely on .... the properties of the ground. .... need the detailed predictions of ground behavior .... ....  routine  .... multiple soil parameters ....

% Multivariatedata (which are commonly encountered in site investigation pro-grams) can be substantially incomplete and typically indirect (that is, a transformation model is needed to convert the measurements tothe desired design parameters .....  which is usally ``incomplete'', i.e., missing values, ..... .... a probability model derived from the target-site data alone may have significant statistical uncertainty \cite{ching2020value} ..... ....A usual practice is to fill the spare target-site data with domain knowledge \cite{sharma2022hierarchical}. ..... .... The retrieval from a databaset of geological or geotechnical data :) 

\section{\shim{Generalized Labeled Multi-Bernoulli Filters}}\label{s:GLMB}
In this section, we introduce a summary of LRFSs and the GLMB filter using the convention \cite{vo2013labeled,vo2014labeled} in a geotechnical viewpoint. \beom{Interested readers are referred to the overview paper \cite{vo2024overview}.} The Kronecker delta $\delta_{Y}[X]$ and the indicator function $1_{Y}(X)$ for arbitrary arguments $X$ and $Y$ have value 1 if $X=Y$ and $X\subseteq Y$ respectively, or 0 otherwise. The class of finite subsets of $X$ is represnted by $\mathcal{F}(X)$. We use the exponential $h^{X}=\prod_{x\in X}h(x)$ of the function $h(x)$, where $h^{\emptyset}=1$. Note that we can omit the current step \bigquotes{$k$} and express the previous step and the next step 
 by \bigquotes{$-$} and \bigquotes{$+$} instead of \bigquotes{$k-1$} and \bigquotes{$k+1$} if there is no ambiguity.

\subsection{\shim{Labeled RFSs}}\label{ss:LRFS}
    \shim{Multiple geotechnical properties can be represented by LRFSs, a finite-set-valued random variable in a Bayesian estimation. Let $\textbf{x} = (x, \ell) \in \mathbb{X} \times \mathbb{L}$ be the multi-property state, where $\mathbb{X}$ represents a state space, and $\mathbb{L}$ is a discrete label space. A starting step $k$ and an identifier $i$ at the same $k$ composes $\ell = (k, i)$, and we also use $\mathcal{L}: \mathbb{X} \times \mathbb{L} \rightarrow \mathbb{L}$ defined by $\mathcal{L}(\ell)$ for the label projection. Moreover, $\mathbb{B}$ denotes the label space for properties \beom{which} are newly measured at $k$ with $\mathbb{L} = \mathbb{L}_{\text{-}} \cup \mathbb{B}$ and $\mathbb{L}_{\text{-}} \cap \mathbb{B} = \emptyset$. The multi-property value are denoted by $\textbf{X} \in \mathcal{F}(\mathbb{X} \times \mathbb{L})$, and the set of distinct labels in $\textbf{X}$ is \beom{expressed} by $\mathcal{L}(\textbf{X})$. We use the label indicator as $\Delta(\textbf{X}) \triangleq \delta_{|\textbf{X}|}\big[|\mathcal{L}(\textbf{X})|\big]$ such that $\Delta(\textbf{X}) = 1$.}
    
\subsection{\shim{GLMB filtering}}\label{ss:GLMB}
    \shim{The LRFS for multi-property are distributed by following the GLMB form, and all information is included in the GLMB filtering density
    \begin{equation}
        \boldsymbol{\pi}(\textbf{X})=\Delta(\textbf{X})\sum_{(I,c)\in\mathcal{F}(\mathbb{L})\times\mathbb{C}}{w^{(I,c)}\delta_{I}[\mathcal{L}(\textbf{X})]\Big[p^{(c)}\Big]^{\textbf{X}}},
    \end{equation} where $\mathbb{C}$ is some finite space, $w^{(I,c)}$ is a non-negative weight, and $p^{(c)}(x,\ell)$ is a probability density. Then, the GLMB density on $\mathcal{F}(\mathbb{X}\times\mathbb{L})$ is simply expressed as 
    \begin{equation}
        \boldsymbol{\pi}\triangleq\Big\{\Big(w^{(I,c)},p^{(c)}\Big)\Big\}_{(I,c)\in\mathcal{F}(\mathbb{L})\times\mathbb{C}},
        \label{eq:glmbrfs}
    \end{equation}
    by using two parameters.}
    
    \shim{Given a current GLMB filtering density by (\ref{eq:glmbrfs}), the GLMB filtering density at next step is 
    \begin{equation}
        \boldsymbol{\pi}_{\text{+}}=\Big\{\Big(w_{Z_{\text{+}}}^{(I_{\text{+}},c,\theta_{\text{+}})},p_{Z_{\text{+}}}^{(c,\theta_{\text{+}})}\Big)\Big\}_{(I_{\text{+}},c,\theta_{\text{+}})\in\mathcal{F}(\mathbb{L}_{\text{+}})\times\mathbb{C}\times\Theta_{\text{+}}},
    \end{equation}
    according to the Bayes recursion. Although the GLMB recursion was originally introduced in separate steps, i.e., \textit{prediction and update}, recently a \textit{GLMB joint prediction and update} \cite{vo2013labeled,vo2014labeled,vo2017efficient,beard2020solution,shim2023linear}, is generally used via 
    \begin{equation}
        \boldsymbol{\pi}_{\text{+}}=\Omega\Big(\boldsymbol{\pi};\boldsymbol{f}_{B}^{(\mathbb{B}_{\text{+}})},Z_{\text{+}}\Big):\mathcal{G}_{\mathbb{L}}\rightarrow\mathcal{G}_{\mathbb{L}_{\text{+}}},
    \end{equation}
    where $\boldsymbol{f}_{B}^{(\mathbb{B}_{\text{+}})}$ is the birth density, $Z_{\text{+}}$ is the set of new measurements, and $\mathcal{G}_{\mathbb{L}}$ is the space of GLMB densities with label space $\mathbb{L}$.}

    \begin{figure*}[b!]
        % \vspace{-0.2cm}
        \centering
    \hspace{-0.5cm}
        \begin{subfigure}[t]{0.3\textwidth}
            \centering
            \includegraphics[height=7.3cm]{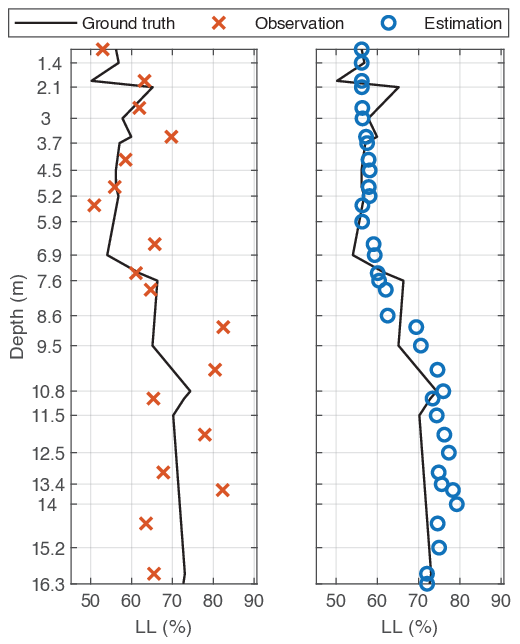}
            \caption{Liquid limit (LL)}
        \end{subfigure}
    \hspace{0.5cm}
        \begin{subfigure}[t]{0.3\textwidth}
            \centering
            \includegraphics[height=7.3cm]{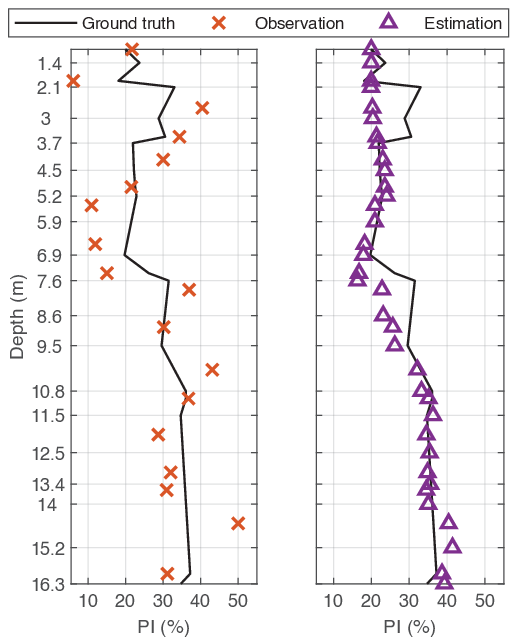}
            \caption{Plasticity index (PI)}
        \end{subfigure}
    \hspace{0.5cm}
        \begin{subfigure}[t]{0.3\textwidth}
            \centering
            \includegraphics[height=7.3cm]{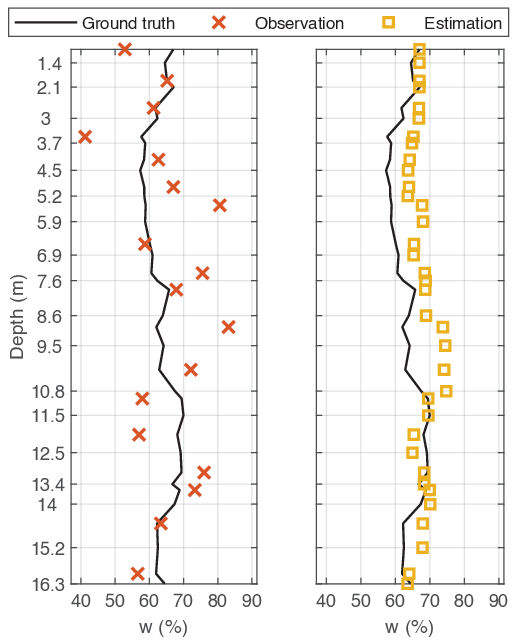}
            \caption{Natural water content (w)}
        \end{subfigure}
        \caption{\shim{Experimental results for clay property estimation on the Ons{\o}y site of Norway.}}
        \label{fig:Norway}
    \end{figure*}

% \ing{The most time-consuming stage in the filter implementation. .... [??]}

\shim{The GLMB filter lends itself to handling BID because the elements of MUSIC-X geotechnical data are well-matched with the characteristics of GLMB filtering. There is also a strong possibility of connection between two different topics, which has been evident through the previous works, e.g., \cite{ching2006bayesian,ching2015statistical,tao2020predicting,lu2020application}. The crux of exploiting LRFS methods in this work is not an argument about finding the best solution for the estimation problem but discussing the feasibility of an established method for geotechnics. Moreover, (relational) database researches in computer science are built on the mathematical set theory like the GLMB filter, which leads researchers and engineers in geotechnical engineering to expect the development of novel multidisciplinary techniques.}

    \begin{table}[t!]
    % \vspace{-0.3cm}
    \renewcommand{\arraystretch}{1.0}
    \caption{\shim{Ons{\o}y site data of Norway.}}
    % \vspace*{-0.1cm}
    \small
        \begin{center}
            % \begin{tabular}{cccccccccccccccccc}
            \begin{tabular}{>{\centering\arraybackslash}p{3mm}>{\centering\arraybackslash}p{15mm}|>{\centering\arraybackslash}p{13mm}>{\centering\arraybackslash}p{13mm}>{\centering\arraybackslash}p{13mm}}
            \toprule
                \textbf{No.} & \textbf{Depth (\textit{m})} &\textbf{LL (\%)} & \textbf{PI (\%)} & \textbf{w (\%)} \\
                % \textbf{No.} & \textbf{Depth (m)} &\textbf{LL ($\boldsymbol{Y_1}$)} & \textbf{PI ($\boldsymbol{Y_2}$)} & \textbf{LI ($\boldsymbol{Y_3}$)} \\
            \midrule
                1 & 1.03 & 56.20 & 19.96 & 66.99 \\
                2 & 1.42 & 56.85 & 23.68 & 64.54 \\
                3 & 1.93 & 50.23 & 18.09 & 65.15 \\
                4 & 2.11 & 65.16 & 33.01 & 67.07 \\
                5 & 2.71 & 60.26 & 30.22 & 61.77 \\
                6 & 3.00 & 57.79  & 28.80 & 62.36 \\
                7 & 3.53 & 59.92 & 30.53 & 57.63 \\
                8 & 3.71 & 57.07 & 21.95 & 58.79 \\
                9 & 4.18 & 56.52 & 22.11 & 58.44 \\
                10 & 4.49 & 56.16 & 22.22 & 57.32 \\
                11 & 4.96 & 56.21 & 22.56 & 58.50 \\
                12 & 5.22 & 56.80 & 22.90 & 58.56 \\
                13 & 5.48 & 56.37 & 22.40 & 58.92 \\
                14 & 5.95 & 55.61 & 21.52 & 58.76 \\
                15 & 6.59 & 54.56 & 20.31 & 60.14 \\
                16 & 6.90 & 54.05 & 19.72 & 60.93 \\
                17 & 7.42 & 61.92 & 26.14 & 60.58 \\
                18 & 7.63 & 66.32 & 31.49 & 62.32 \\
                19 & 7.89 & 66.16 & 31.23 & 65.77 \\
                20 & 8.63 & 65.68 & 30.46 & 63.91 \\
                21 & 8.95 & 65.46 & 30.13 & 62.03 \\
                22 & 9.48 & 65.12 & 29.58 & 64.17 \\
                23 & 10.17 & 70.02 & 33.02 & 62.88 \\
                24 & 10.78 & 74.37 & 36.07 & 67.50 \\
                25 & 10.99 & 72.86 & 36.07 & 69.43 \\
                26 & 11.48 & 70.18 & 34.69 & 69.97 \\
                27 & 12.02 & 70.53 & 35.00 & 68.17 \\
                28 & 12.54 & 70.85 & 35.28 & 69.16 \\
                29 & 13.10 & 71.21 & 35.60 & 69.39 \\
                30 & 13.44 & 71.42 & 35.79 & 66.74 \\
                31 & 13.60 & 71.53 & 35.88 & 68.85 \\
                32 & 14.00 & 71.78 & 36.10 & 67.35 \\
                33 & 14.56 & 72.13 & 36.41 & 62.24 \\
                34 & 15.25 & 72.56 & 36.80 & 62.48 \\
                35 & 15.99 & 73.03 & 37.21 & 61.96 \\
                36 & 16.28 & 72.67 & 34.73 & 64.46 \\
                % 1 & 1.03 & 56.20 & 19.96 & 1.54 \\
                % 2 & 1.42 & 56.85 & 23.68 & 1.32 \\
                % 3 & 1.93 & 50.23 & 18.09 & 1.82 \\
                % 4 & 2.11 & 65.16 & 33.01 & 1.06 \\
                % 5 & 2.71 & 60.26 & 30.22 & 1.05 \\
                % 6 & 3.00 & 57.79  & 28.80 & 1.16 \\
                % 7 & 3.53 & 59.92 & 30.53 & 0.93 \\
                % 8 & 3.71 & 57.07 & 21.95 & 1.08 \\
                % 9 & 4.18 & 56.52 & 22.11 & 1.09 \\
                % 10 & 4.49 & 56.16 & 22.22 & 1.05 \\
                % 11 & 4.96 & 56.21 & 22.56 & 1.10 \\
                % 12 & 5.22 & 56.80 & 22.90 & 1.07 \\
                % 13 & 5.48 & 56.37 & 22.40 & 1.11 \\
                % 14 & 5.95 & 55.61 & 21.52 & 1.15 \\
                % 15 & 6.59 & 54.56 & 20.31 & 1.27 \\
                % 16 & 6.90 & 54.05 & 19.72 & 1.35 \\
                % 17 & 7.42 & 61.92 & 26.14 & 0.95 \\
                % 18 & 7.63 & 66.32 & 31.49 & 0.87 \\
                % 19 & 7.89 & 66.16 & 31.23 & 0.99 \\
                % 20 & 8.63 & 65.68 & 30.46 & 0.94 \\
                % 21 & 8.95 & 65.46 & 30.13 & 0.89 \\
                % 22 & 9.48 & 65.12 & 29.58 & 0.97 \\
                % 23 & 10.17 & 70.02 & 33.02 & 0.78 \\
                % 24 & 10.78 & 74.37 & 36.07 & 0.81 \\
                % 25 & 10.99 & 72.86 & 36.07 & 0.90 \\
                % 26 & 11.48 & 70.18 & 34.69 & 0.99 \\
                % 27 & 12.02 & 70.53 & 35.00 & 0.93 \\
                % 28 & 12.54 & 70.85 & 35.28 & 0.95 \\
                % 29 & 13.10 & 71.21 & 35.60 & 0.95 \\
                % 30 & 13.44 & 71.42 & 35.79 & 0.87 \\
                % 31 & 13.60 & 71.53 & 35.88 & 0.93 \\
                % 32 & 14.00 & 71.78 & 36.10 & 0.88 \\
                % 33 & 14.56 & 72.13 & 36.41 & 0.73 \\
                % 34 & 15.25 & 72.56 & 36.80 & 0.73 \\
                % 35 & 15.99 & 73.03 & 37.21 & 0.70 \\
                % 36 & 16.28 & 72.67 & 34.73 & 0.76 \\
            \bottomrule
        \end{tabular}
        \label{tbl:Norway}
    \end{center}
    \vspace{-1.5cm}
    \end{table}

\section{\shim{Numerical Study}}\label{s:exp}
\shim{This section shows experimental results by utilizing GLMB filtering on BID. We use the clay database, CLAY/10/7490, to evaluate the performance of proposed approach, especially with LL, PI, and the water content (w) of each site in Norway and Taiwan, respectively. For our experiments, measured data in the database are regarded as ``(data) \textit{ground truths}'' because it is difficult to obtain the actual value of a physical quantity \cite{sharma2023spectral}. Thus, we remove some raw data, i.e., ground truths,  and noises are also added on the raw data for generating incomplete noisy ``\textit{observations}''. A general linear Gaussian setup is used for filtering. Note that we are interested in not analyzing specific geotechnical databases but estimating clay properties from MUSIC-X sensor data with LRFS approaches as accurate as possible.}

    \begin{figure*}[b!]
    % \vspace{-0.2cm}
        \centering
    \hspace{-0.5cm}
        \begin{subfigure}[t]{0.3\textwidth}
            \centering
            \includegraphics[height=7.3cm]{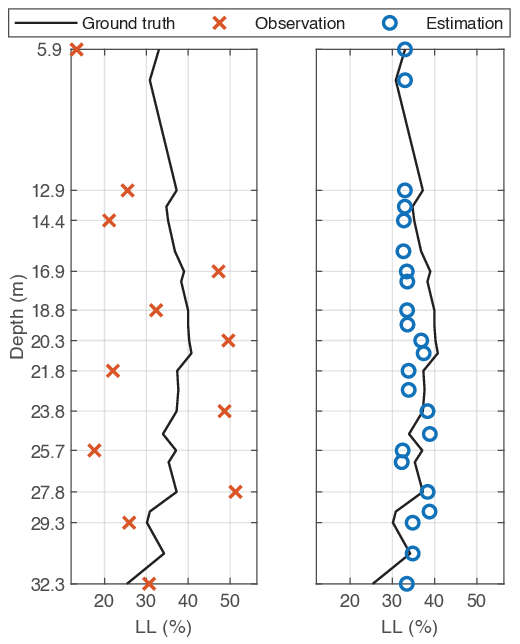}
            \caption{Liquid limit (LI)}
        \end{subfigure}
    \hspace{0.5cm}
        \begin{subfigure}[t]{0.3\textwidth}
            \centering
            \includegraphics[height=7.3cm]{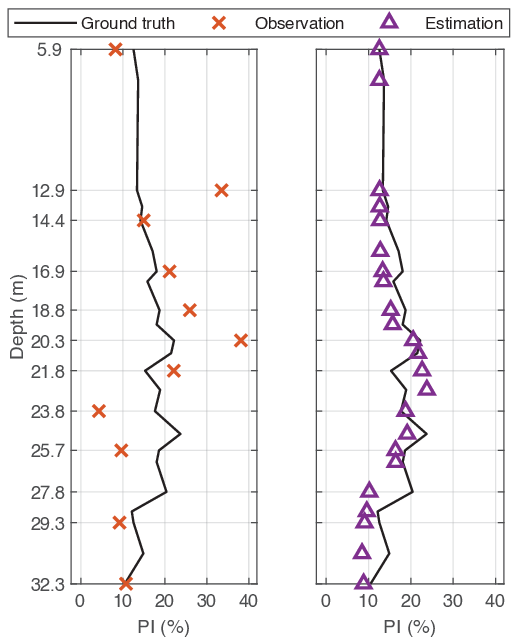}
            \caption{Plasticity index (PI)}
        \end{subfigure}
    \hspace{0.5cm}
        \begin{subfigure}[t]{0.3\textwidth}
            \centering
            \includegraphics[height=7.3cm]{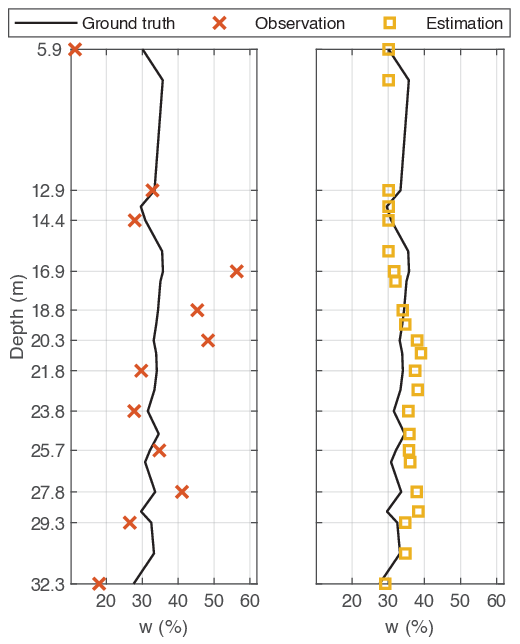}
            \caption{Natural water content (w)}
        \end{subfigure}
        \caption{\shim{Experimental results for clay property estimation on the Taipei site of Taiwan.}}
        \label{fig:Taiwan}
    \end{figure*}

\subsection{\shim{Case study 1: Ons{\o}y site, Norway}}\label{subsec:Norway}
    \shim{The first study is based on Ons{\o}y data at 36 different depths from 1.03\:$m$ to 16.28\:$m$. All values of three clay properties at each depth are listed in Table~\ref{tbl:Norway}. For the filter implementation, a single data point is composed of a 2D vector $[x, \dot{x}]^{\textsf{T}}$ of a 1D sensor value and deviation between consecutive values, and the sampling interval varies by matching with the ground truth for simplicity. Although we can measure clay properties continuously in practice, sensors usually miss some information, which depends on sensor performance. We use a detection probability of $P_D = 0.5$ to control the sparsity of imperfect observations, which means 50 noisy observations among 100 true values are available. The observation noise standard deviation is $\sigma_m = 10\:\%$. The initial values are assumed to be roughly known as engineers can generally consider existing data in similar sites. Process noise standard deviation is $\sigma_p = 0.3\:\%$ per interval squared for all properties.}
    % $\sigma_p = 0.3\:\%/m^{2}$ for all properties.}

    \shim{Fig.~\ref{fig:Norway} shows estimation results via filtering on the above Ons{\o}y observations. Ground truths (raw data) for LL are respectively shown as a black line graph with generated observations (orange cross) and estimates (blue circle) in Fig.~\ref{fig:Norway}(a). Due to noisy and incomplete sensor readings, it is not easy to obtain correct patterns from observations. However, estimated patterns generally follow the patterns of ground truths. Missing values are also recovered with other further corrected values. Fig.~\ref{fig:Norway}(b) shows the result of PI estimation (purple triangle) with the same setting including observations (orange cross). The pattern of PI in the ground truth similarly fluctuates with that of LL. The proposed approach shows better property values than observations despite less accuracy compared to the case of LL. Furthermore, experiments for w and resultant estimates (yellow square) are presented in Fig.~\ref{fig:Norway}(c). It is observed that the trend of w is quite linear than those of two other properties. Notwithstanding relatively noticeable difference between ground truths and observations, our estimation generally captures its patterns. Consequently, the proposed methodology is expected to fit not only MOT but also geotechnical applications by providing additional accurate data (estimated) and then enabling researchers and engineers to scrutinize BID.}

    \begin{table}[t!]
    % \vspace{-0.3cm}
    \renewcommand{\arraystretch}{1.0}
    \caption{\shim{Taipei site data of Taiwan.}}
    % \vspace*{-0.1cm}
    \small
        \begin{center}
            % \begin{tabular}{cccccccccccccccccc}
            \begin{tabular}{>{\centering\arraybackslash}p{3mm}>{\centering\arraybackslash}p{15mm}|>{\centering\arraybackslash}p{13mm}>{\centering\arraybackslash}p{13mm}>{\centering\arraybackslash}p{13mm}}
            % \begin{tabular}{m{0.3cm}m{0.3cm}|p{0.45cm}|p{0.45cm}|p{0.45cm}|p{0.45cm}|p{0.45cm}|p{0.45cm}|p{0.45cm}|p{0.45cm}}
            \toprule
                \textbf{No.} & \textbf{Depth (\textit{m})} &\textbf{LL (\%)} & \textbf{PI (\%)} & \textbf{w (\%)} \\
                % \textbf{No.} & \textbf{Depth (m)} & \textbf{LL ($\boldsymbol{Y_1}$)} & \textbf{PI ($\boldsymbol{Y_2}$)} & \textbf{LI ($\boldsymbol{Y_3}$)} \\
            \midrule
                1 & 5.89 & 33.00 & 12.56 & 30.10 \\
                2 & 7.42 & 30.83 & 13.66 & 35.70 \\
                3 & 12.87 & 37.23 & 13.38 & 33.40 \\
                4 & 13.67 & 34.78 & 14.62 & 29.58 \\
                5 & 14.35 & 35.18 & 14.21 & 30.85 \\
                6 & 15.88 & 36.82 & 17.10 & 35.52 \\
                7 & 16.88 & 39.00 & 18.07 & 35.76 \\
                8 & 17.38 & 38.32 & 15.86 & 35.04 \\
                9 & 18.79 & 39.95 & 18.76 & 34.38 \\
                10 & 19.50 & 39.95 & 18.07 & 33.87 \\
                11 & 20.29 & 40.23 & 22.20 & 33.17 \\
                12 & 20.92 & 40.77 & 21.51 & 33.86 \\
                13 & 21.79 & 37.36 & 15.31 & 34.04 \\
                14 & 22.74 & 37.64 & 18.89 & 33.38 \\
                15 & 23.79 & 37.23 & 17.65 & 31.53 \\
                16 & 24.92 & 33.96 & 23.72 & 34.58 \\
                17 & 25.73 & 37.09 & 18.62 & 32.11 \\
                18 & 26.31 & 35.32 & 18.07 & 30.77 \\
                19 & 27.78 & 37.23 & 20.41 & 33.59 \\
                20 & 28.75 & 30.83 & 12.14 & 29.65 \\
                21 & 29.30 & 30.14 & 12.56 & 32.50 \\
                22 & 30.83 & 34.23 & 14.90 & 33.25 \\
                23 & 32.33 & 25.38 & 10.49 & 27.58 \\
                % 1 & 5.89 & 33.00 & 12.56 & 0.77 \\
                % 2 & 7.42 & 30.83 & 13.66 & 1.36 \\
                % 3 & 12.87 & 37.23 & 13.38 & 0.71 \\
                % 4 & 13.67 & 34.78 & 14.62 & 0.64 \\
                % 5 & 14.35 & 35.18 & 14.21 & 0.69 \\
                % 6 & 15.88 & 36.82 & 17.10 & 0.92 \\
                % 7 & 16.88 & 39.00 & 18.07 & 0.82 \\
                % 8 & 17.38 & 38.32 & 15.86 & 0.79 \\
                % 9 & 18.79 & 39.95 & 18.76 & 0.70 \\
                % 10 & 19.50 & 39.95 & 18.07 & 0.66 \\
                % 11 & 20.29 & 40.23 & 22.20 & 0.68 \\
                % 12 & 20.92 & 40.77 & 21.51 & 0.68 \\
                % 13 & 21.79 & 37.36 & 15.31 & 0.78 \\
                % 14 & 22.74 & 37.64 & 18.89 & 0.77 \\
                % 15 & 23.79 & 37.23 & 17.65 & 0.68 \\
                % 16 & 24.92 & 33.96 & 23.72 & 1.03 \\
                % 17 & 25.73 & 37.09 & 18.62 & 0.73 \\
                % 18 & 26.31 & 35.32 & 18.07 & 0.75 \\
                % 19 & 27.78 & 37.23 & 20.41 & 0.82 \\
                % 20 & 28.75 & 30.83 & 12.14 & 0.90 \\
                % 21 & 29.30 & 30.14 & 12.56 & 1.19 \\
                % 22 & 30.83 & 34.23 & 14.90 & 0.93 \\
                % 23 & 32.33 & 25.38 & 10.49 & 1.21 \\
            \bottomrule
        \end{tabular}
        \label{tbl:Taiwan}
    \end{center}
    \vspace{-1.0cm}
    \end{table}

\subsection{\shim{Case study 2: Taipei site, Taiwan}}\label{subsec:Taiwan}
    \shim{The next test is conducted on Taipei site data, measured at 23 depths from 5.89\,$m$ to 32.33\,$m$. Table~\ref{tbl:Taiwan} presents the sensor readings in an increasing depth order. Unlike the previous case, the gap between consecutive observations usually exceeds 1\:$m$ despite total observed data are smaller than that of Ons{\o}y site. We use the same setting with the first case study for the comparison purpose and its results are shown in Fig.~\ref{fig:Taiwan}. Specifically, Fig.~\ref{fig:Taiwan}(a) presents the results of GLMB filtering for clay property, LL, estimation. The observation uncertainty level is similar to experiments on Ons{\o}y site data, but estimated clay parameters are relatively less accurate. This is mainly because of fewer observations given the same depth interval. Furthermore, the case of PI resembles that of LL, shown in Fig.~\ref{fig:Taiwan}(b). Conversely but also as expected, our approach shows similar performance for w estimation on both sites, see Fig.~\ref{fig:Taiwan}(c). It can be seen that the implementation of GLMB filtering has a high level of ability to filter out noise and interpolate missing points, which can lead a wide range of geotechnical applications.}

\section{\shim{Conclusion}}\label{s:conclusion}
    \shim{This paper has studied clay property estimation in the geotechnical database using  Bayesian estimation with LRFSs. We have particularly utilized the GLMB filter which has been popularized in the MOT literature along with considerable potential for a host of applications. In this work, the distribution of clay properties is generated through GLMB filtering by regarding multiple properties as multi-object states in the MOT problem. Then, the estimates are finally obtained by maximum a posteriori, analogous to trajectory estimation. The performance of our approach has been evaluated by a series of experiments based on the real geotechnical database, CLAY/10/7490. In Norway and Taiwan cases, clay property estimation is achieved from MUSIC-X measurements without geotechnical knowledge, whereupon we anticipate different strong solutions for site retrieval problems in both academia and industry.}
    
    \shim{Future works include in-depth study on theoretical connection between LRFS methods and geotechnical databases because this work is limited as the feasibility study irrelevant to mathematical discussion or proofs. Given further detailed works, various LRFS approaches, i.e, established MOT solutions including smoothing \cite{moratuwage2022multi}, are expected as promising techniques for dealing with the site recognition challenge. Scalable methodologies also have been requested in the data-centric geotechnics community \cite{yang2020scalable,indraratna2021large}. Moreover, visual analytics of high-dimensional data \cite{kim2021skyflow,kim2022three} can be another prospective direction due to the need for user-friendly tools handling multivariate geotechnical data.}

\bibliographystyle{IEEEtran}
\balance
\bibliography{IEEEabrv, GPE_ref}

\end{document}